\documentclass[conference]{IEEEtran}
\IEEEoverridecommandlockouts
\usepackage{cite}
\usepackage{amsmath,amssymb,amsfonts}
\usepackage{algorithmic}
\usepackage{graphicx}
\usepackage{textcomp}
\usepackage{xcolor}
\usepackage{tikz}
\usepackage{orcidlink}
\usepackage{subcaption}
\usepackage{siunitx}
\usepackage{mwe}
\usepackage[ruled,vlined]{algorithm2e}
\usepackage{comment}
\usepackage{tikz}
\usepackage{hyperref}
\usepackage[capitalise]{cleveref}
\usepackage{pgfplots}
\usepackage{cite}
\usepackage{amsmath,amssymb,amsfonts}
\usepackage{algorithmic}
\usepackage{graphicx}
\usepackage{textcomp}
\usepackage{wrapfig}
\usepackage{booktabs}
\usepackage{multirow}
\usepackage{tabularx}

\usepackage{pifont}

\SetKwComment{Comment}{$\triangleright$\ }{}
\usetikzlibrary{mindmap,shadows}
\usetikzlibrary{arrows.meta}
\usetikzlibrary{positioning}

\usetikzlibrary{arrows.meta, mindmap, shadows}
\usetikzlibrary{arrows}
\usetikzlibrary{backgrounds}

\pgfdeclarelayer{bg}    
\pgfsetlayers{bg,main}  

\newcommand{\RM}[1]{\MakeUppercase{\romannumeral #1{}}}

\newcommand{\p}[1]{{\text{\Pisymbol{psy}{#1}}}}

\def\BibTeX{{\rm B\kern-.05em{\sc i\kern-.025em b}\kern-.08em
    T\kern-.1667em\lower.7ex\hbox{E}\kern-.125emX}}

\begin{document}

\title{Auto-Positioning in Radio-based Localization Systems: A Bayesian Approach\\

}



\author{\IEEEauthorblockN{1\textsuperscript{st} Andrea Jung\orcidlink{0000-0003-1019-6134}}
\IEEEauthorblockA{\textit{Institute of Traffic Telematics} \\
\textit{Technische Universit\"at Dresden}\\
Dresden, Germany \\
andrea.jung@tu-dresden.de}
\and
\IEEEauthorblockN{2\textsuperscript{nd} Paul Schwarzbach\orcidlink{0000-0002-1091-782X}}
\IEEEauthorblockA{\textit{Institute of Traffic Telematics} \\
\textit{Technische Universit\"at Dresden}\\
Dresden, Germany \\
paul.schwarzbach@tu-dresden.de}
\and
\IEEEauthorblockN{3\textsuperscript{rd} Oliver Michler\orcidlink{0000-0002-8599-5304}}
\IEEEauthorblockA{\textit{Institute of Traffic Telematics} \\
\textit{Technische Universit\"at Dresden)}\\
Dresden, Germany \\
oliver.michler@tu-dresden.de}
}

\maketitle

\begin{abstract}
The application of radio-based positioning systems is ever increasing. In light of the dissemination of the Internet of Things and location-aware communication systems, the demands on localization architectures and amount of possible use cases steadily increases. While traditional radio-based localization is performed by utilizing stationary nodes, whose positions are absolutely referenced, collaborative auto-positioning methods aim to estimate location information without any a-priori knowledge of the node distribution. The usage of auto-positioning decreases the installation efforts of localization systems and therefore allows their market-wide dissemination. Since observations and position information in this scenario are correlated, the uncertainties of all nodes need to be considered. In this paper we propose a discrete Bayesian method based on a multi-dimensional histogram filter to solve the task of robust auto-positioning, allowing to propagate historical positions and estimated position uncertainties, as well as lowering the demands on observation availability when compared to conventional closed-form approaches. The proposed method is validated utilizing different multipath-, outlier and failure-corrupted ranging measurements in a static environment, where we obtain at least 58\% higher positioning accuracy compared to a baseline closed-form auto-positioning approach.
\end{abstract}


\section{Introduction}
\label{sec:intro}

The development of location-based services (LBS) enabled by radio-based localization comprises a vast majority of indoor positioning systems (IPS) \cite{Laoudias_network_localization_survey_2018}. With the on-going integration of communication and localization systems \cite{Zhang_enabling_jcas_survey_2022}, especially in the context of the Internet of Things (IoT) \cite{Shafique_iot_5g_narrow_band_survey_2020} and future, beyond 5G mobile communication systems \cite{bourdoux_6g_2020-3}, diversification of conventional radio-based IPS is constantly increasing. This includes technologies \cite{MendozaSilva2019metareviewips}, network architectures \cite{Zhang_enabling_jcas_survey_2022}, use cases and corresponding positioning scenarios \cite{Li_assisted_living_wireless_sensing_isac_6G_2022}. 

Traditionally, radio-based localization is performed by classifying network nodes into two categories: stationary anchors or base stations, whose positions are known, and mobile tags, whose locations are of interest \cite{Zafari2019IPSSurvey}. With this in mind, the localization of mobile nodes is only achievable when a certain amount of stationary devices are present and their locations is precisely determined a-priori. However, the aforementioned diversification and on-going network densification possibly leads to a rise in IPS at a scale, where anchor-individual surveying will not be feasible anymore. 

\begin{figure}[!htbp]
    \centering
    \vspace{.2cm}
    \includegraphics[width=0.38\textwidth]{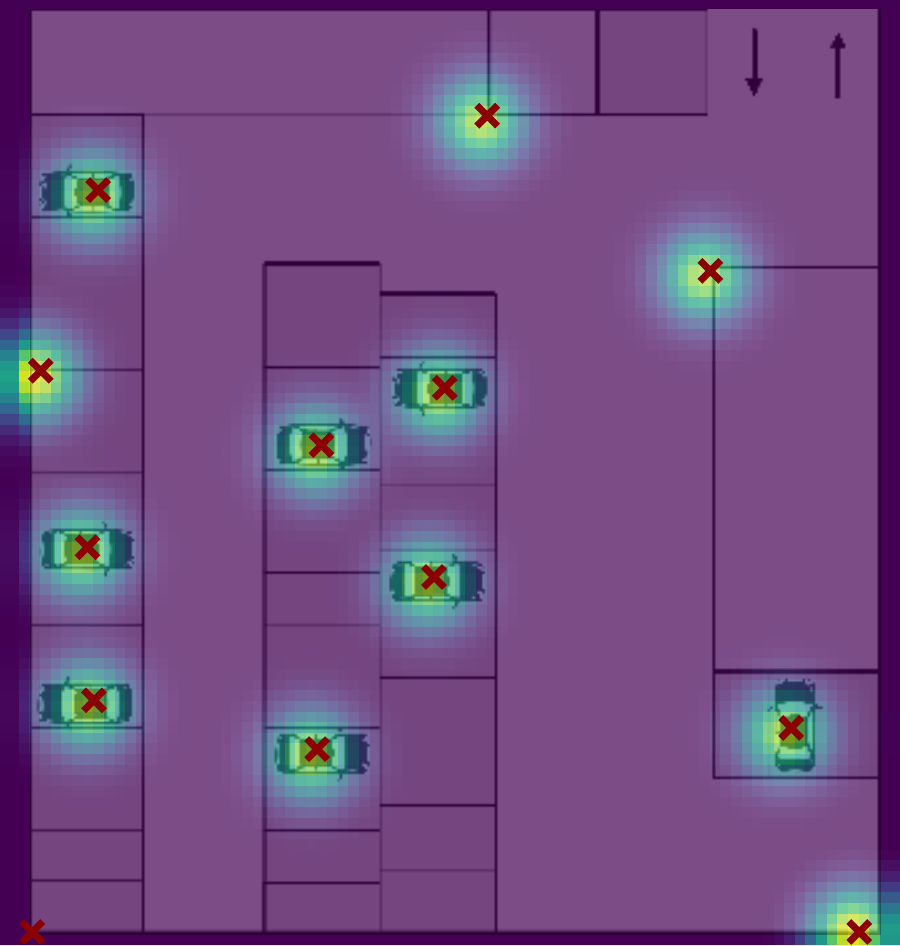}  
    \caption{Collaborative auto-positioning for smart parking applications: State estimation and associated uncertainties based on probability grid mapping (purple, reference positions in red).}
    \label{fig:Future Parking Application}
\end{figure}

In addition, a strict distinction between stationary and mobile devices will not be applicable anymore due to the variant nature of many applications, where variability of the environment is a main focus and non-stationary configurations are imperatively required. While this may be regarded challenging conceptionally, these new architectures also provide the capabilities for immersive IPS and ehance their scalability. 

A potential use case in the context of intelligent transportation systems (ITS) is depicted in \cref{fig:Future Parking Application} \cite{jung_wsn_future_parking_2020}, where LBS for in-house parking are provided. In this example, the amount of stationary anchors can drastically be reduced by incorporating quasi-static devices representing parking cars. This use case is addressed as efficient parking is one of the main challenges for individual motorized transportation in urban areas \cite{ninnemannMultipathassistedRadioSensing2022}. In this context, the usage of auto-positioning enables a time- and cost-efficient roll out of IPS by dispensing positional surveying of stationary nodes. In addition, static infrastructure can be reduced as quasi-static nodes can be incorporated. This also corresponds to a variety of use cases in other application fields.

\subsection{Status quo of Auto-Positioning}
\label{ssec:status_quo}

For this task, collaborative auto-positioning, also known as self-calibration, is intended as the automatic procedure that allows anchor nodes to  identify their own locations without additional interventions, e.g. manual surveying \cite{Ridolfi_self_calibration__overview_2021}. In the context of location-aware communication systems, auto-positioning can aid to increase the applicability and scalability of both device-based and device-free localization \cite{Liu_fundamental_limits_integrated_sensing_2022}. This holds true, especially for dense, location-aware networks like 5G \cite{Hakkarainen_5g_localization_dense_networks_2015, Kim_5g_mmwave_cooperative_positioning_2020}.

The basic idea of auto-positioning is related to collaborative localization, which has been intensively studied in the past years \cite{ Chukhno_d2d_cooperative_positioning_survey_2022}, where inter-mobiles ranges are used to support the localization process and still rely on previously surveyed locations of the reference points. A comprehensive survey and literature review of self-calibration and collaborative localization, especially emphasizing their differences and similarities is given in \cite{Ridolfi_self_calibration__overview_2021}. Theoretical works with regards to achievable performance are provided in \cite{HamerD'Andrea2018}, where the Cramér-Rao lower bound for auto-positioning is investigated. 

A common approach to auto-positioning without any pre-surveyed anchor positions is the usage of closed-form (CF) methods as proposed in \cite{HamerD'Andrea2018, Almansa_autocalib_multi_robot_2020, Rapid_deploy_2021}. These methods are based on similar assumptions for auto-positioning, which will further be discussed in \cref{sec:closed-form}. However, CF methods require the simultaneous incorporation of observations obtained from multiple nodes. Hence, the success rate in the presence of unknown constellations and measurement failures is limited. 

A Least-Squares-Estimation (LSE) approach is used in \cite{Almansa_autocalib_multi_robot_2020} to minimize the errors between the inter-anchor measurements. The calibration of the anchors is done if a certain error threshold is exceeded. The automate coordinate formation in \cite{Rapid_deploy_2021} is supplemented with a node placement strategy and an outlier removal algorithm. Ref.~\cite{anchor_calib_real_time_2020} utilizes additional calibration modules within the network in order to reduce the positional error. For this additional anchor calibration, the performance of three localization algorithms is tested via multidimensional scaling, semidefinite programming and iterative trilateration. 

In order to further increase the accuracy of auto-positioning based on trilateration by means of hardware and ranging enhancements, \cite{Extendable_Autopositioning_2021} proposes an antenna delay calibration based on an Asymmetric Double-sided Two-way Ranging (ADS-TWR) scheme. Furthermore, \cite{Pereira_phd_self_calib_2021} proposes two novel algorithms to improve the accuracy and success rate of auto-positioning, namely Triangle Reconstruction Algorithm (TRA) and Channel Impulse Response Positioning (CIRPos). Both algorithms, were tested in a simulated environment. With regards to technologies, most of the cited works investigated the proposed methods based on Ultra-Wideband (UWB) respectively simulation procedures.

In general, auto-positioning leads to correlated inter-node observations, whose uncertainties with respect to the position information and measurement noise need to be considered during estimation. Especially in the presence of non-line-of-sight and multipath propagation robust state estimation is required, as these error types lead to non-gaussian residual distributions, which hurt the presumptions of estimators like the LSE. The works in \cite{Ridolfi_self_calib_NLOS_2021, anchor_calib_real_time_2020, Extendable_Autopositioning_2021, Pereira_phd_self_calib_2021} target the accuracy improvement of self-calibration by identifying non-line-of-sight observations, e.g. by applying machine learning. 

\subsection{Focus and structure of this paper}
\label{ssec:focus}
This paper presents a grid-based Bayesian formulation of the collaborative auto-positioning problem for IPS. A grid-based representation was chosen in order to provide a shared state space for collaborative users and to potentially include a-priori knowledge about the environment. The presented, non-parametric filtering approach provides a robust state estimation compared to conventional CF methods for non-stationary and unknown network configurations, while lowering the requirements with regards to connectivity and availability of viable ranging measurements. To underline this, we use multipath- and outlier-corrupted simulation data aiming to provide a real-world proximate data foundation for method validation. The simulation procedure emulates UWB range measurements with respect to three different scenarios. 

The rest of the paper is organized as follows: \cref{sec:closed-form} describes a baseline CF auto-positioning method. The there described relations are used for initialization of the grid-based auto-positioning method, which is presented in \cref{sec:approach}. In order to validate the proposed method, a brief introduction on the applied empirical simulation method for three different ranging residual distributions and quantitative positioning accuracy results for these scenarios are given in \cref{sec:results}. The paper concludes with a summary and proposals for future research work in \cref{sec:conclusion}.

\section{Closed-Form Auto-Positioning}
\label{sec:closed-form}

The aforementioned CF methods for auto-positioning based on distance measurements are applied by meeting a variety of presumptions. In order to estimate the positions of three anchors $\boldsymbol{A}_0$, $\boldsymbol{A}_1$ and $\boldsymbol{A}_2$ within a network, the methods proposed in \cite{HamerD'Andrea2018} and \cite{Pereira_phd_self_calib_2021} formulate the following presumptions: 

\begin{itemize}
\item $\boldsymbol{A}_0$ is situated at the coordinate origin;
\item The direction from $\boldsymbol{A}_0$ to $\boldsymbol{A}_1$ defines the positive x-axis;
\item $\boldsymbol{A}_2$ lies in the half-plane with positive y-coordinate;
\item Extension: $\boldsymbol{A}_3$ lies in the positive z-direction.
\end{itemize}

\begin{figure}[ht]
    \centering
    \includegraphics[width=0.9\linewidth, trim={1.0cm 0.8cm 1.0cm 0.8cm},clip]{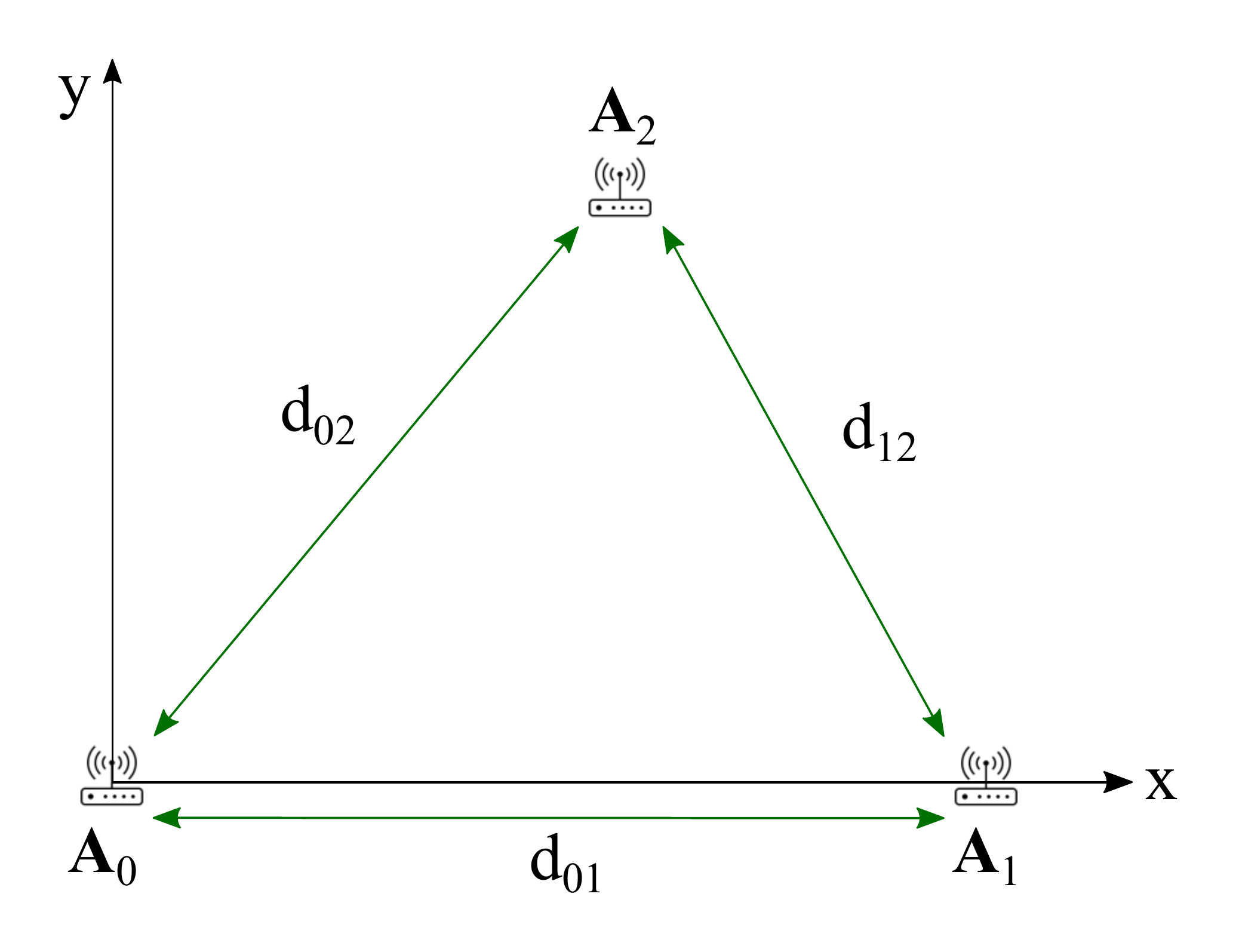}
    \caption{Configuration of network nodes $\boldsymbol{A}_0$ to $\boldsymbol{A}_2$ and their pair-wise distances $d$.}
    \label{fig:autopositioning_method}
\end{figure}

A corresponding two-dimensional constellation and the provided pair-wise distance measurements $\boldsymbol{d}$ are depicted in \cref{fig:autopositioning_method}. Given this frame, each anchor position $\boldsymbol{A}_n$ is defined as $\boldsymbol{A}_n = [x_n, y_n]^\intercal$. Assuming a total of $N$-nodes, the inter-anchor distances form the square measurement matrix $\boldsymbol{D}_t$ at timestep $t$:

\begin{equation*}
\boldsymbol{D}_{t} = 
\begin{pmatrix}
d_{0,0} & d_{0,1} & \cdots & d_{0,N} \\
d_{1,0} & d_{1,1} & \cdots & d_{1,N} \\
\vdots  & \vdots  & \ddots & \vdots  \\
d_{n,0} & d_{N,1} & \cdots & d_{N,N} 
\end{pmatrix}
\end{equation*}

Please note, that the x-coordinate of $\boldsymbol{A}_1$ represents the measured distance between itself and node $\boldsymbol{A}_0$.

\begin{align}
    \boldsymbol{A}_0 &= [ x_{0}, y_{0} ]^\intercal = [ 0, 0 ]^\intercal \label{eq:Anchor0} \\
    \boldsymbol{A}_1 &= [ x_{1}, y_{1} ]^\intercal = [ d_{01}, 0 ]^\intercal  \label{eq:Anchor1} \\
    \boldsymbol{A}_2 &= [ x_{2}, y_{2} ]^\intercal = [ \frac{d_{02}^2 - d_{12}^2 + x_{1}^2}{2x_{1}}, \sqrt{d_{02}^2 - x_{2}^2}]^\intercal  \label{eq:Anchor2}
\end{align}

Additional nodes in the network can be calculated via \cref{eq:Anchor2} or estimated based on the position data of the first three nodes and their ranges to them. A detailed derivation of \cref{eq:Anchor2} is given in \cite{Rapid_deploy_2021}. 

Given CF approaches, we identify two major challenges when it comes to real-world applications: Presence of corrupted measurements (multipath reception and outliers) and measurement failures. Concerning the latter, both the pair-wise distances $d_{0, 1}$ between  $\boldsymbol{A}_0$ and $\boldsymbol{A}_1$ and the respective ranging measurements to the node of interest $n$ $d_{0, n}$ and $d_{1, n}$ need to be available in order to estimate node positions for $n > 1$. In the presence of measurement confusions, failures and nodes possibly being out of reception range, this leads to descending success rates, which we will further discuss in \cref{sec:results}. In addition, CF methods as well as parametric estimators are performing poorly in the presence of non-line-of-sight (NLOS) and multipath reception as well as correlation between observations. Therefore real-world scenarios provide inherent challenges for these methods, which need to be considered. 

\section{Bayesian Auto-Positioning Approach}
\label{sec:approach}

In order to do so, we propose a Bayesian formulation based on a discrete grid in order to solve the auto-positioning problem. This Markov Localization recursively estimates the a-posteriori probability density function (pdf), commonly reffered to as belief or posterior, of the current state via the observation Likelihood, while incorporating process knowledge and state history via the Markov assumption and Bayes' rule in order to provide a more robust state estimation \cite{thrun2005probabilistic}:

\begin{equation}
    \text{posterior} \propto \text{likelihood} \times \text{prior}
    \label{eq:bayes_in_words}
\end{equation}

\subsection{Fundamentals}
\label{subsec:probability grid positioning}
The proposed method, which we will refer to as collaborative grid positioning (CGP) is based on an equidistant grid representation of the state space, which therefore represents possible realizations of the node's location. This representation was chosen because it provides a shared and unified state space of multiple nodes while dispensing a foundation for non-parametric state estimation, which is more robust towards non-gaussian measurement noise and multi-modalities. The method is also referred to as multi-dimensional histogram filter (HF) \cite{thrun2005probabilistic}, which corresponds to the point-mass filter \cite{Bucy_digital_synthesis_non_linear_filters_1971}.

For this method, the state space given the two-dimensional state vector $\mathcal{X}_t = [x_t, y_t]^\intercal$ representing the position of a node, is decomposed in a discrete and finite set of $M$-equidistant realizations $X_M$:

\begin{equation}
    \text{dom}(\boldsymbol{\mathcal{X}})= \boldsymbol{X}_1 \cup \boldsymbol{X}_2 \cup \dots \boldsymbol{X}_M
    \label{equ:discrete_domain}
\end{equation}

\begin{figure}[ht]
    \centering
    \scalebox{.7}{\tikzset{%
  >={Latex[width=2mm,length=2mm]},
            base/.style = {rectangle, rounded corners, draw=black,
                           minimum width=2.5cm, minimum height=1.3cm,
                           text width = 2.5cm, fill=gray!20,
                           text centered},
  			standard/.style = {rectangle, rounded corners, draw=black,
                           minimum width=2.5cm, minimum height=1.3cm,
                           text width = 2.5cm, fill=gray!20,
                           text centered},
            line/.style={draw, very thick, ->},
            standard1/.style = {rectangle, rounded corners, draw=black,
                           minimum width=2.5cm, minimum height=1.3cm,
                           text width = 2.5cm, fill=gray!20,
                           text centered},
}
   
\begin{tikzpicture}[node distance=1.9cm,
    every node/.style={fill=white}, align=center]
  \node (start)             [standard]              {{\textbf{Initialization} \\ $\text{dom}(\boldsymbol{\mathcal{X}})$}};
  \node (predict) [base, above of=start, yshift=.6cm] {{\textbf{Prediction} \\ $\overline{\boldsymbol{p}}_{t}$}};
  \node (observe) [base, right of=predict, xshift=2.5cm] {{\textbf{Observation}\\$\boldsymbol{p}_{t}$}};
  
  \node (data) [standard] at (observe |- start) {{\textbf{Observations} \\ $\boldsymbol{M}_t$}};
  
  \begin{pgfonlayer}{bg} 
   \begin{scope}
        \node (data2) [standard, xshift=0.26cm, yshift=.26cm] at (observe |- start) {};
        \node (data1) [standard, xshift=0.13cm, yshift=0.13cm] at (observe |- start) {};
       \end{scope}
 \end{pgfonlayer}

  \node (estimate) [standard1, right of=observe, xshift=1.75cm] {{\textbf{Estimation}\\$\boldsymbol{\hat{\mathcal{X}}_t}$}};

   \path[line] (start) -- node[midway,left, xshift=-0.1cm] {{$t=0$}} (predict);
   \path[line] (data) -- node[midway,right, xshift=0.1cm] {{$t=1:T$}} (observe);
   \path[line] (observe) -- (estimate);
    \path[line, <-] (predict) to[bend right=50](observe);
    \path[line] (predict) to[bend left=50](observe);
  \end{tikzpicture}}
    \caption{Conceptual structure of CGP.}
    \label{fig:grid_flowchart}
\end{figure}
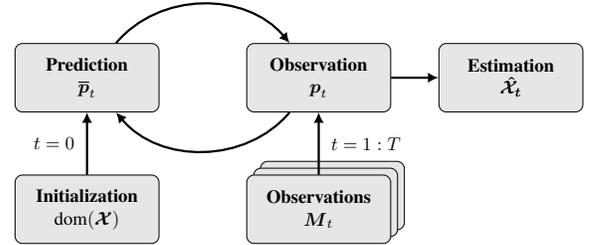

The general procedure for the estimation of a single node, following the well-known Recursive Bayes' Filter structure, is given in \cref{fig:grid_flowchart}, where the hidden state space vector $\boldsymbol{\mathcal{X}}_{t}$ is incrementally estimated based on the last 
given state $\boldsymbol{\mathcal{X}}_{t-1}$. 

The corresponding calculations are given as follows \cite{thrun2005probabilistic}:

\begin{align}
    \overline{\boldsymbol{p}}_{t} &= \boldsymbol{p}_{t-1} \sum_m \boldsymbol{P}(\boldsymbol{\mathcal{X}}_{t, m}|{\boldsymbol{\mathcal{X}}}_{t-1}) \\
    \boldsymbol{p}_{t} &= \eta\; \overline{\boldsymbol{p}}_{t} \sum_m \boldsymbol{P} (\mathcal{\boldsymbol{\mathcal{Z}}}_{t} |\boldsymbol{\mathcal{X}}_{t, m}) \\
    \boldsymbol{\hat{\mathcal{X}}}_{t} &= \text{argmax}\;\boldsymbol{p}_{t},
    \label{equ:notation}
\end{align}

where $\overline{\boldsymbol{p}}_{t}$ and $\boldsymbol{p}_{t}$ denote the predicted and the resulting belief, based on the observation Likelihood $\boldsymbol{P}(\mathcal{\boldsymbol{\mathcal{Z}}}_{t}|\boldsymbol{\mathcal{X}}_{t, m})$ and the normalization constant~$\eta$. The resulting state estimation $\boldsymbol{\hat{\mathcal{X}}}_{t}$ is obtained from maximizing the current belief. 

\subsection{Inter-node Likelihood calculation}
\label{ssec:likelihood}
In contrast to conventional radio-based localization systems, where observations are only obtained from stationary anchors, collaborative auto-positioning needs to consider and propagate the associated uncertainties about the locations of the network nodes. This also poses a major challenge in collaborative positioning scenarios \cite{Minetto_cognitive_collaborative_positioning_2020}.

\begin{figure}[htb!]
    \centering
    \resizebox{.85\linewidth}{!}{

\begin{tikzpicture}[node distance=5em,
    base/.style = {rectangle,draw=black, thick,
                           minimum width=12mm, minimum height=12mm, fill=black!10,
                           text width = 11mm,
                           text centered},
    rounded/.style = {circle, rounded corners, draw=black, thick,  fill=black!5,
                           minimum width=12mm, minimum height=3mm,
                           text width = 11mm,
                           text centered},
    dots/.style = {rectangle, rounded corners, draw=white,
                           minimum width=5mm, minimum height=15mm,
                           text width = 27mm,
                           text centered}
    every node/.style={fill=white, font=\sffamily}, align=center]

  \node (x_k) [rounded] {$\boldsymbol{\mathcal{X}}_k$};
  \node (x_k1) [rounded, right of=x_k, xshift=10mm] {$\boldsymbol{\mathcal{X}}_{k+1}$};
  \node (x_k2) [rounded, right of=x_k1, xshift=10mm] {$\boldsymbol{\mathcal{X}}_{k+2}$};
  \node (x_kn) [rounded, right of=x_k2, xshift=10mm] {$\boldsymbol{\mathcal{X}}_{k+n}$};
  
  \node (z_k) [base, below of=x_k, yshift=-9mm] {$\boldsymbol{\mathcal{Z}}_{k}$};
  \node (z_k1) [base, below of=x_k1, yshift=-9mm] {$\boldsymbol{\mathcal{Z}}_{k+1}$};
  \node (z_k2) [base, below of=x_k2, yshift=-9mm] {$\boldsymbol{\mathcal{Z}}_{k+2}$};
  \node (z_kn) [base, below of=x_kn, yshift=-9mm] {$\boldsymbol{\mathcal{Z}}_{k+n}$};
  
  \node (y_k) [rounded, below of=z_k, yshift=-9mm] {$\boldsymbol{\mathcal{X}}_k$};
  \node (y_k1) [rounded,  below of=z_k1, yshift=-9mm] {$\boldsymbol{\mathcal{X}}_{k+1}$};
  \node (y_k2) [rounded, below of=z_k2, yshift=-9mm] {$\boldsymbol{\mathcal{X}}_{k+2}$};
  \node (y_kn) [rounded, below of=z_kn, yshift=-9mm] {$\boldsymbol{\mathcal{X}}_{k+n}$};
  
  \draw[black, -triangle 90] (x_k) -- node[midway, right, xshift=1mm] {$\boldsymbol{P} (\mathcal{\boldsymbol{\mathcal{Z}}}_{t} |\boldsymbol{\mathcal{X}}_{t})$} (z_k);
  \draw[black, -triangle 90] (x_k1) -- node[midway, above, yshift=1mm] {} (z_k1);
  \draw[black, -triangle 90] (x_k2) -- node[midway, above, yshift=1mm] {} (z_k2);
  \draw[black, -triangle 90] (x_kn) -- node[midway, above, yshift=1mm] {} (z_kn);
  
    \draw[black, -triangle 90] (y_k) -- node[midway, right, xshift=1mm] {$\boldsymbol{P} (\mathcal{\boldsymbol{\mathcal{Z}}}_{t} |\boldsymbol{\mathcal{X}}_{t})$} (z_k);
  \draw[black, -triangle 90] (y_k1) -- node[midway, above, yshift=1mm] {} (z_k1);
  \draw[black, -triangle 90] (y_k2) -- node[midway, above, yshift=1mm] {} (z_k2);
  \draw[black, -triangle 90] (y_kn) -- node[midway, above, yshift=1mm] {} (z_kn);
  
  \draw[black, -triangle 90] (x_k) to[bend left] node[midway, above] {$\boldsymbol{P}(\boldsymbol{\mathcal{X}}_{t+1}|{\boldsymbol{\mathcal{X}}}_{t})$} (x_k1);
  \draw[black, -triangle 90] (x_k1) to[bend left] node[midway, above, yshift=3mm] {} (x_k2);
  \draw[black, -triangle 90] (x_k2) to[bend left] node[midway, above, yshift=1mm] {\Huge{...}} (x_kn);
  
  \draw[black, -triangle 90] (y_k) to[bend right] node[midway, below] {$\boldsymbol{P}(\boldsymbol{\mathcal{X}}_{t+1}|{\boldsymbol{\mathcal{X}}}_{t})$} (y_k1);
  \draw[black, -triangle 90] (y_k1) to[bend right] node[midway, below, yshift=-3mm] {} (y_k2);
  \draw[black, -triangle 90] (y_k2) to[bend right] node[midway, below, yshift=-1mm] {\Huge{...}} (y_kn);
  
  \node[text width=3cm] at (10.5, 0) {Node 1};
  \node[text width=3cm] at (10.5, -2.7) {Observations};
  \node[text width=3cm] at (10.5, -5.3) {Node 2};
  
 
  \end{tikzpicture}
  }
    \caption{Graphical representation of the state estimation problem based on correlated inter-node observations.}
    \label{fig:state_graphical_model}
\end{figure}
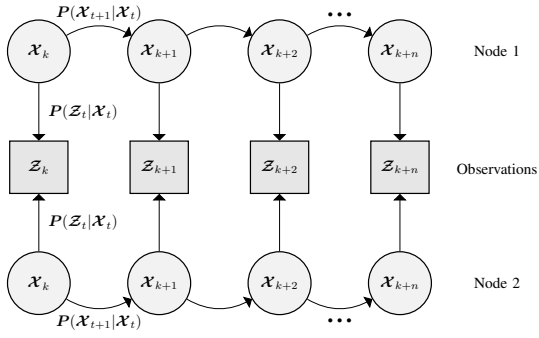

In order to underline this problem, the correlation and statistical dependency is depicted in \cref{fig:state_graphical_model}, which shows the correlation dependencies of inter-node observations.

\vspace{-.5cm}

Therefore, the uncertainty of both the ranging measurements $\sigma_r$ as well as the uncertainties of the node estimations $\sigma_{\boldsymbol{A}_n}$ need to be accounted for Likelihood calculation. Given two nodes $\boldsymbol{A}_{1,2}$ and in correspondence to \cite{Medina_hybrid_estimators_2020}, the inter-node ranging noise is given as:

\begin{equation}
    \sigma_{1,2} = \sigma_r + \sqrt{\textrm{tr}(\hat{\boldsymbol{\Sigma}}_{1})} + \sqrt{\textrm{tr}(\hat{\boldsymbol{\Sigma}}_{2})},
    \label{equ:variance_propagation}
\end{equation}

where $\textrm{tr}(\cdot)$ denotes the trace operator. The node-individual covariance matrices $\boldsymbol{\Sigma}_n$ can be computed as the sample variance given the previously defined state space and the calculated Likelihood for each sample by determining the weighted average:

\begin{equation}
    \boldsymbol{\Sigma}_{n,t} \approx \sum_m \boldsymbol{p}_m (\boldsymbol{X}_m - \hat{\boldsymbol{\mathcal{X}}}_{n,t})^2
\end{equation}

\begin{figure}[htb!]
    \centering
    \begin{subfigure}[b]{0.17\textwidth}
    \centering
    \includegraphics[width=\linewidth]{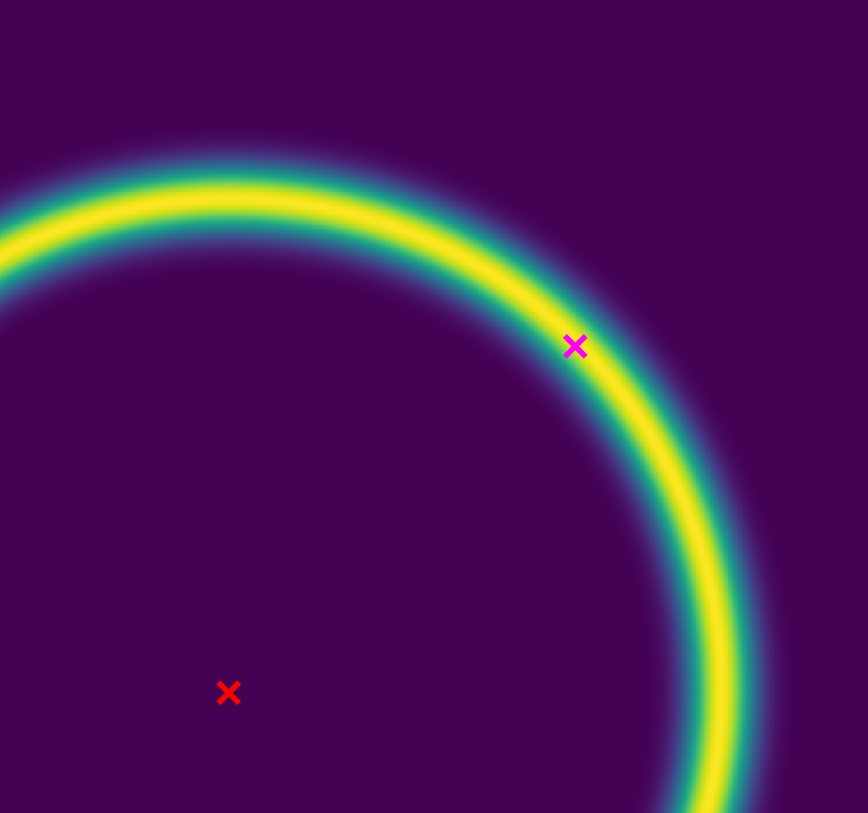}
    \caption{Stationary}
    \label{fig:grid_var_range}
    \end{subfigure}
    \hspace{1cm}
    \centering
    \begin{subfigure}[b]{0.17\textwidth}
    \centering
    \includegraphics[width=\linewidth]{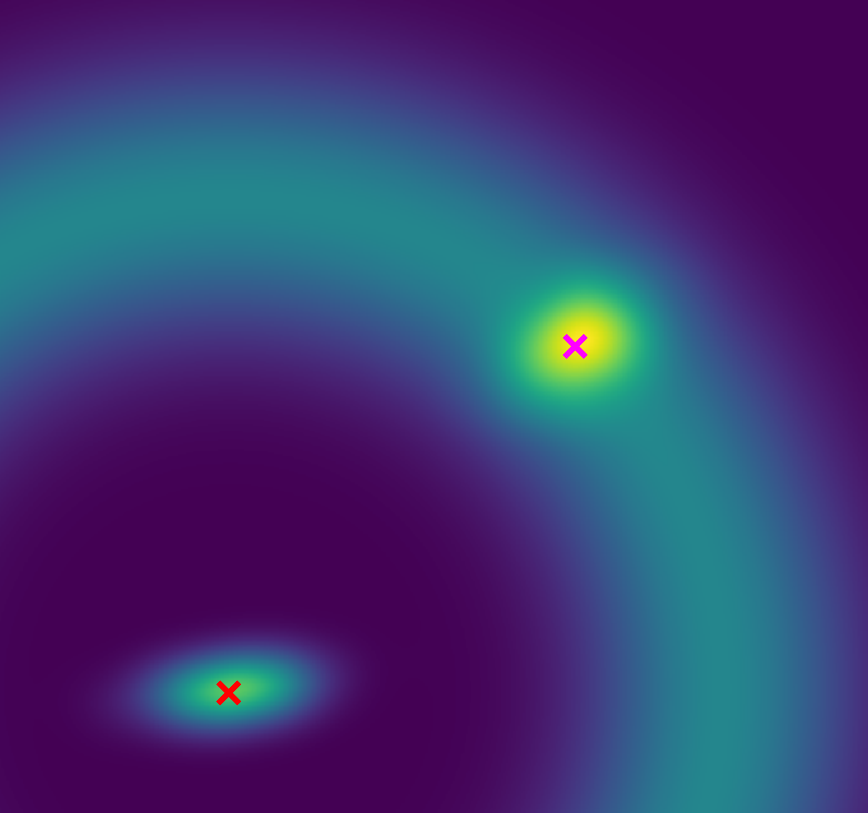}
    \caption{Non-stationary}
    \label{fig:grid_var_prop}
    \end{subfigure}
    \caption{Observation variance: \textbf{(a)} Uncorrelated case with stationary nodes; \textbf{(b)} Correlated case for non-stationary nodes.}
    \label{fig:grid_variance}
\end{figure}

A graphical example of the measurement uncertainties between a conventional range measurement obtained from a stationary anchor compared to auto-positioning is given in \cref{fig:grid_variance}.

Given the auto-positioning problem at hand, the observation Likelihood based on the ranging measurement $r$ can be sampled from a normal distribution:

\begin{equation}
    \boldsymbol{P}(\mathcal{\boldsymbol{\mathcal{Z}}}|\boldsymbol{\mathcal{X}}_m)  \leftarrow  \mathcal{N}(\boldsymbol{y}_{m}, \boldsymbol{\Sigma})
    \label{equ:normal_sampling}
\end{equation}

where $y$ denotes the euclidean distance residual between the observed measurement and $m$-grid-node relation: 

\begin{equation}
    \boldsymbol{y}_m = \left \|\boldsymbol{X}_n - \boldsymbol{X}_m\right \|_2 - \boldsymbol{r} 
    \label{equ:markov_residual}
\end{equation}

\subsection{Implementational Details}
\begin{figure}[ht]
    \centering
    \includegraphics[width=0.48\textwidth, trim={0.0cm 0.0cm 0.0cm 0.0cm},clip]{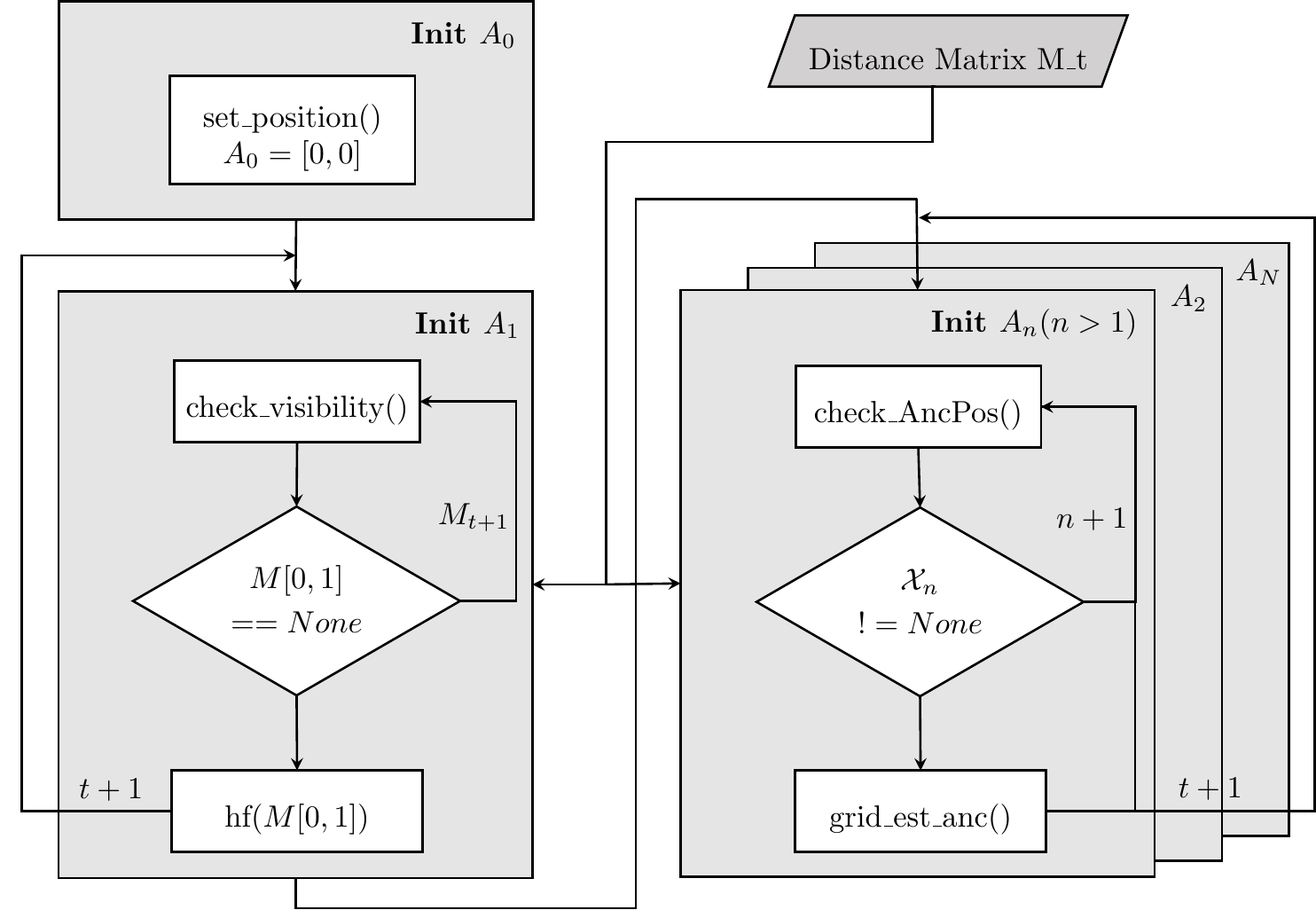}  
    \caption{Flowchart of the proposed CGP auto-positioning.}
    \label{fig:flowchart_approach}
\end{figure}

The structure and a flowchart of the CGP implementation are visualized in \cref{fig:flowchart_approach}. Similar to the CF, the position of $\boldsymbol{A}_0$ is set to the origin of the coordinate system. Next up, the position of $\boldsymbol{A}_1$ is estimated with the proposed CGP approach in a one-dimensional representation, depending on the availability of the ranging measurement to $\boldsymbol{A}_0$. This dimensional reduction corresponds to a HF and can be applied based on the presumptions formulated in \cref{sec:closed-form}.

In contrast to the CF, for CGP the next step is to estimate the remaining sensor node positions based on all previously estimated sensor node positions and their measured distances to the node of interest. This is done by applying the two-dimensional CPG as previously described. In order to lower the demands on the availability of observations compared to the CF approach, CGP does not require specific ranging combinations to be available (cf. \cref{sec:closed-form}), which raises its robustness against measurement failures. This effect will further be discussed in \cref{sec:results}. Therefore, in each estimation step, the visibility to all remaining nodes is checked and available observations in combination with the originating state estimation and associated uncertainty are taken into account.

Due to the non-parametric nature of the CGP, NLOS reception and measurement outliers are compensated to a certain extent, without the need for additional error identification and mitigation, which helps in providing an easily applicable auto-positioning approach.

\section{Validation and Results}
\label{sec:results}
In order to assess the proposed auto-positioning method with respect to the aforementioned challenges, a semi-empirical ranging simulation procedure based on our previously published work \cite{Schwarzbach_statistical_evaluation_uwb_ips_2022} is applied. Since many high-precision positioning systems for IPS are based on the UWB technology, the simulation intends to rebuild typical error types in magnitudes with respect to IPS scenarios. The simulation procedure allows an adaptive tuning of parameters to ensure real-world proximate performance validation. The scenario of choice corresponds to the constellation depiced in \cref{fig:Future Parking Application}.

\subsection{Methodology}
We assume that a ranging measurement $r$ comprises the true, euclidean distance between two nodes $d=\left\| \boldsymbol{A}_1- \boldsymbol{A}_2\right\|_2$ and additive errors $\p{101}$ following \cite{qi2006analysis}:

\begin{align}
	r=\begin{cases} d+\p{101} & \text{if }p>\dfrac{d}{d_\mathrm{max}}\\ \emptyset & \text{else.}\end{cases}
	\label{equ:GenDaten1}
\end{align}

The additive error terms are modeled as linearly distant-dependent with respect to an empirical maximum range $d_\text{max}$, which can be set with regards to the application and technology at hand. If the formulated condition is not met, a measurement failure is simulated. In addition, the probability $p$ for sporadic measurement perturbations is modeled as a Bernoulli experiment with $p \sim \mathcal{U}(0, 1)$. The distance dependency of errors was empirically shown in \cite{Schwarzbach_statistical_evaluation_uwb_ips_2022} and also influences the success rate of simulated measurements. Depending on this ratio, the classification of the error variable is obtained from: 

\begin{align}
	\p{101}=\begin{cases} \p{101}_\mathrm{mp} & \text{if }p_\p{101}>0.8-0.3\dfrac{d}{d_\mathrm{max}},\p{101}_\mathrm{mp}<d\\ \p{101}_\mathrm{out} & \text{if }p_\p{101}<p_\mathrm{out}\\ \p{101}_\mathrm{hw} & \text{else,}\end{cases}
	\label{equ:GenDaten2}
\end{align}

where the outlier probability $p_\text{out}$ is also an empirical value describing the outlier probability. Based on this classification and among the aforementioned measurement failures, three types of errors are sampled:

\begin{align}
    \p{101}_\mathrm{mp}&\sim\mathcal{LN}(\overline{R}_\mathrm{mp},\sigma_\mathrm{mp}^2) \\
    \p{101}_\mathrm{out}&\sim\mathcal U(-d,d_\mathrm{max}-d) \\
    \p{101}_\mathrm{hw}&\sim \mathcal{N}(0,\sigma_r^2).
\end{align}

$\p{101}_\mathrm{hw}$ represent normally distributed LOS measurements, $\p{101}_\mathrm{out}$ a uniformly distributed outlier magnitude with respect to the given reference distance. Finally, $\p{101}_\mathrm{mp}$ is modeled as a right-skewed log normal distribution \cite{Prorok2011_lognorm}, where the skewness of the log normal distribution depends on the diversity of the NLOS channel. The applied parameters and probabilities are summarized in \cref{tab:sim_and_results}.

\subsection{Ranging Simulation}
\begin{figure*}[htb!]
    \centering
    \begin{subfigure}[b]{0.32\textwidth}
    \centering
    \includegraphics[width=\linewidth, trim=.3cm 0cm 1cm 1cm, clip]{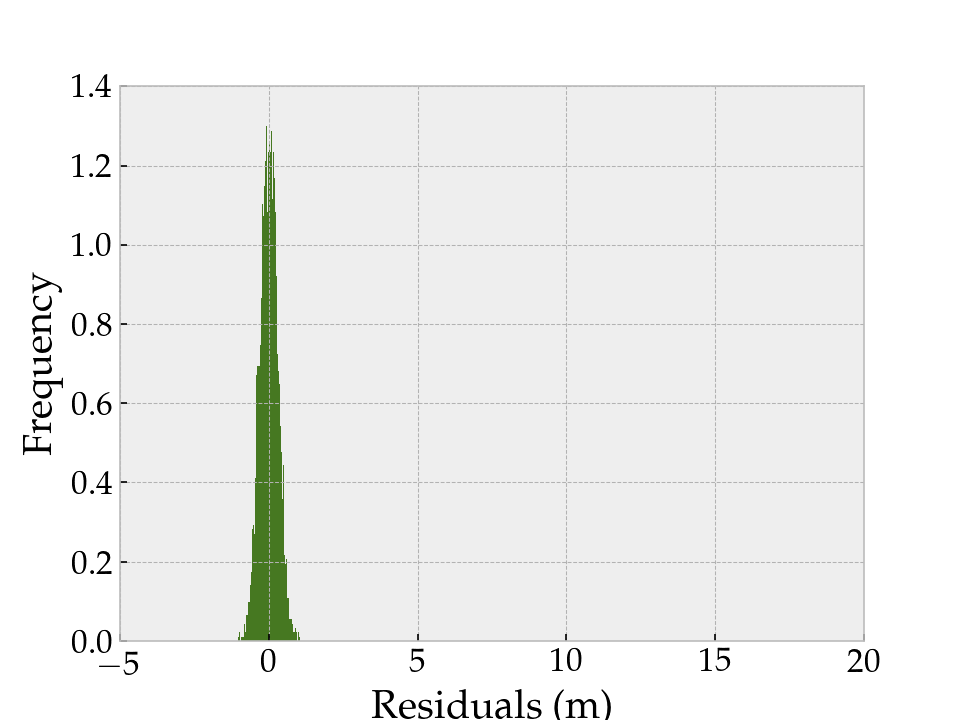}
    \caption{\RM{1}}
    \label{fig:hist0}
    \end{subfigure}
    \centering
    \begin{subfigure}[b]{0.32\textwidth}
    \centering
    \includegraphics[width=\linewidth, trim=.3cm 0cm 1cm 1cm, clip]{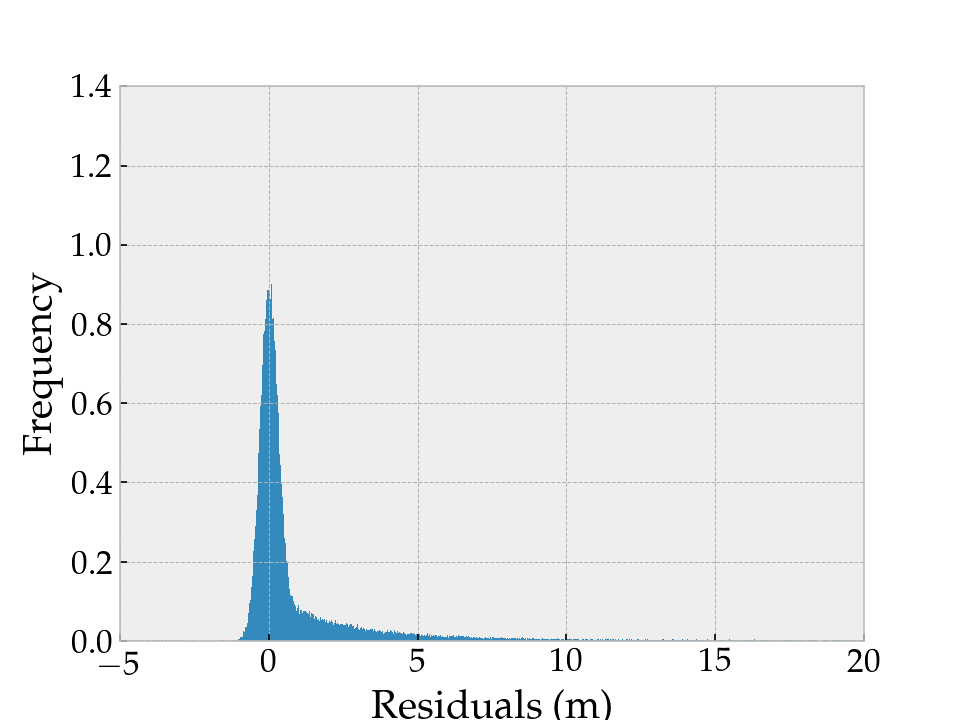}
    \caption{\RM{2}}
    \label{fig:hist1}
    \end{subfigure}
    \centering
    \begin{subfigure}[b]{0.32\textwidth}
    \centering
    \includegraphics[width=\linewidth, trim=.3cm 0cm 1cm 1cm, clip]{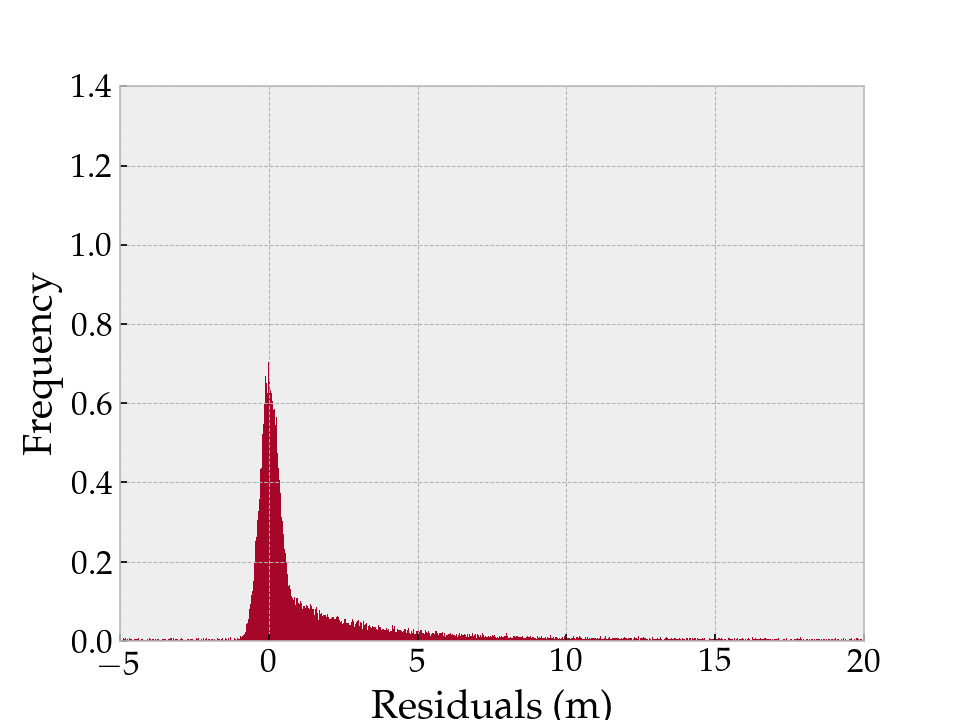}
    \caption{\RM{3}}
    \label{fig:hist2}
    \end{subfigure}
    \caption{Histogram of ranging residual PDFs for the examined scenarios. Simulation parameters are included in \cref{tab:sim_and_results}.}
    \label{fig:residual_hist}
\end{figure*}

In total, three different ranging distributions are simulated, where the measurement residuals for each scenario are shown in \cref{fig:residual_hist} and each scenario emulates different environmental conditions. In addition, \cref{tab:sim_and_results} also contains the relative amount of LOS, NLOS, outlier and failure rates. 

Given a constellation of 13 nodes (cf. \cref{fig:Future Parking Application}), each scenario contains 1.000 measurement epochs and therefore approximately 13.000 estimated positions based on around 170.000 ranging measurements, allowing a statistical assessment of each constellation.

\begin{table}[htbp!]
\caption{Overview of simulation parameterization for three simulation scenarios, percentage ranging simulation results and quantitative positioning performance of both CF and CGP.}
\label{tab:sim_and_results}
\centering

\begin{tabular}{@{}p{1.2cm}p{1.8cm}p{1.cm}p{1.cm}p{1cm}@{}}
\toprule[1.5pt]
                  &   & \RM{1} & \RM{2} & \RM{3} \\ \midrule[1.5pt]
\multirow{5}{*}{\textbf{Parameter}} & PDFs &\multicolumn{3}{c}{$\mathcal{N}(0,0.9) \quad \mathcal{LN}(0.8, 1.07)$} \\ \cmidrule(l){2-5} 
                  & $d_\textrm{max}$ (m)& 100 & 100 & 50 \\ \cmidrule(l){2-5} 
                  & $p_\textrm{out}$ & 0 & 0.07  & 0.20 \\ \cmidrule(l){2-5} 
                  & NLOS (\%)& 0 & 0.21 & 0.20 \\ \midrule[1.5pt]             
\multirow{5}{*}{\textbf{Simulation}} & LOS (\%)& 1 & 0.61  & 0.56 \\ \cmidrule(l){2-5} 
                  & NLOS (\%)& 0 & 0.21 & 0.20 \\ \cmidrule(l){2-5} 
                  & Outlier (\%)& 0 & 0.06 & 0.15 \\ \cmidrule(l){2-5} 
                  & Failures (\%)& 0 & 0.12 & 0.09 \\ \midrule[1.5pt]
\multirow{6}{*}{\textbf{CF}} & RMSE (m)&  0.41 & 3.80 & 6.99 \\ \cmidrule(l){2-5} 
                  & 1-$\sigma$ (m)&  0.51 & 2.06 & 4.01 \\ \cmidrule(l){2-5} 
                  & 2-$\sigma$ (m)&  0.93 & 19.41 & 37.25 \\ \cmidrule(l){2-5} 
                  & 3-$\sigma$ (m)&  1.25 & 47.22 & 62.93 \\ \cmidrule(l){2-5} 
                  & Succes (\%)& 1 & 0.37 & 0.36 \\ \midrule[1.5pt]
\multirow{6}{*}{\textbf{CGP}} &  RMSE (m)& 0.17 & 0.42 & 0.93 \\ \cmidrule(l){2-5} 
                  & 1-$\sigma$ (m)& 0.28 & 0.50 & 1.43 \\ \cmidrule(l){2-5} 
                  & 2-$\sigma$ (m)& 0.36 & 0.58 & 1.57 \\ \cmidrule(l){2-5} 
                  & 3-$\sigma$ (m)& 0.44 & 0.59 & 1.58 \\ \cmidrule(l){2-5} 
                  & Succes (\%)& 1 & 1 &  1\\\bottomrule[1.5pt]                  
\end{tabular}
\end{table}

Scenario \RM{1} corresponds to exclusively gaussian noise representing only LOS measurements. Additionally, scenario \RM{2} incorporates multipath errors, outliers and measurement failures. Finally, scenario \RM{3} puts even more emphasis on occurring outliers.

\begin{figure*}[h!]
    \centering
    \begin{subfigure}[b]{0.32\textwidth}
    \centering
    \includegraphics[width=\linewidth, trim=11.5cm .8cm 12cm 2cm, clip]{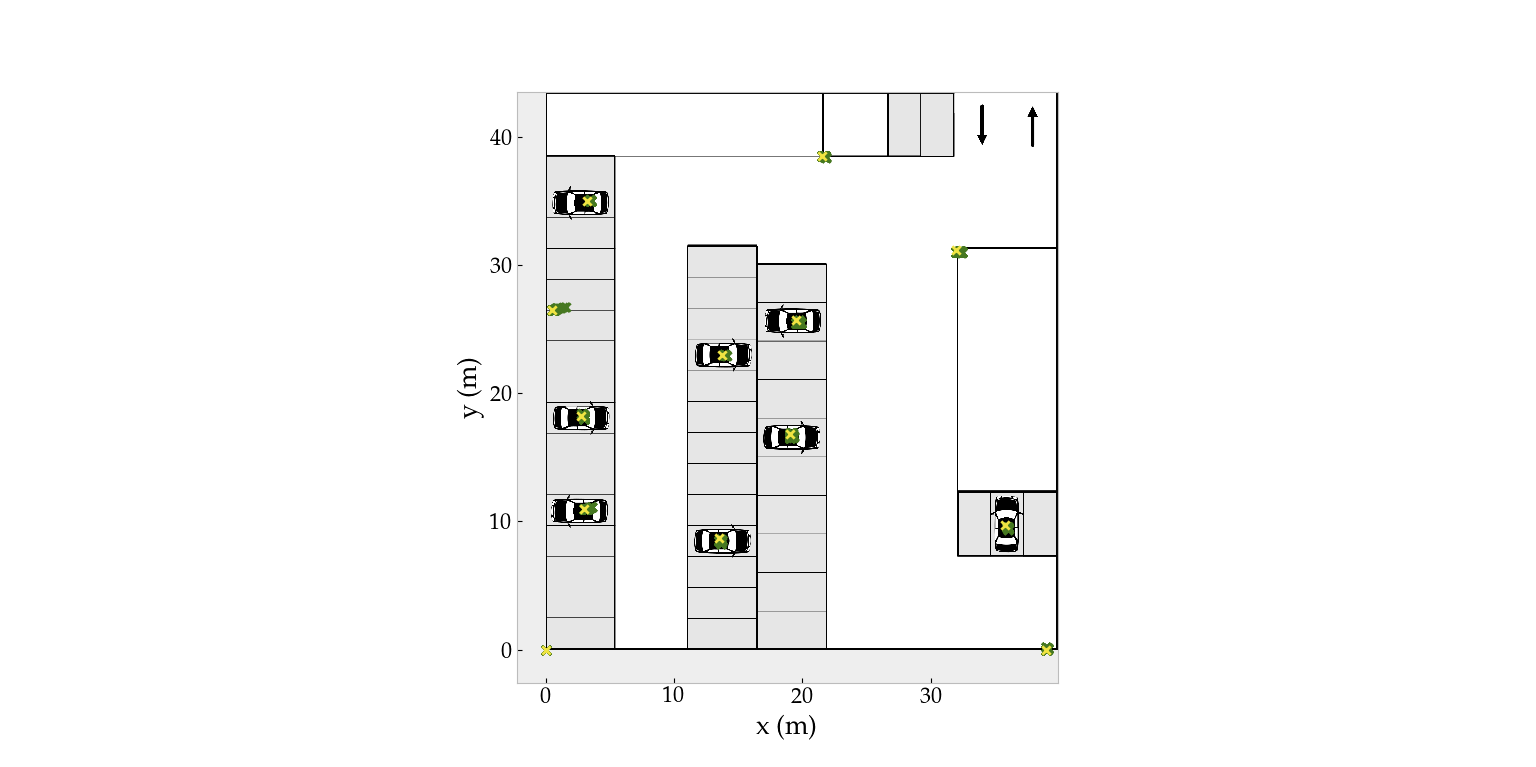}
    \caption{\RM{1}}
    \label{fig:grid_data_green}
    \end{subfigure}
    \centering
    \begin{subfigure}[b]{0.32\textwidth}
    \centering
    \includegraphics[width=\linewidth, trim=11.5cm .8cm 12cm 2cm, clip]{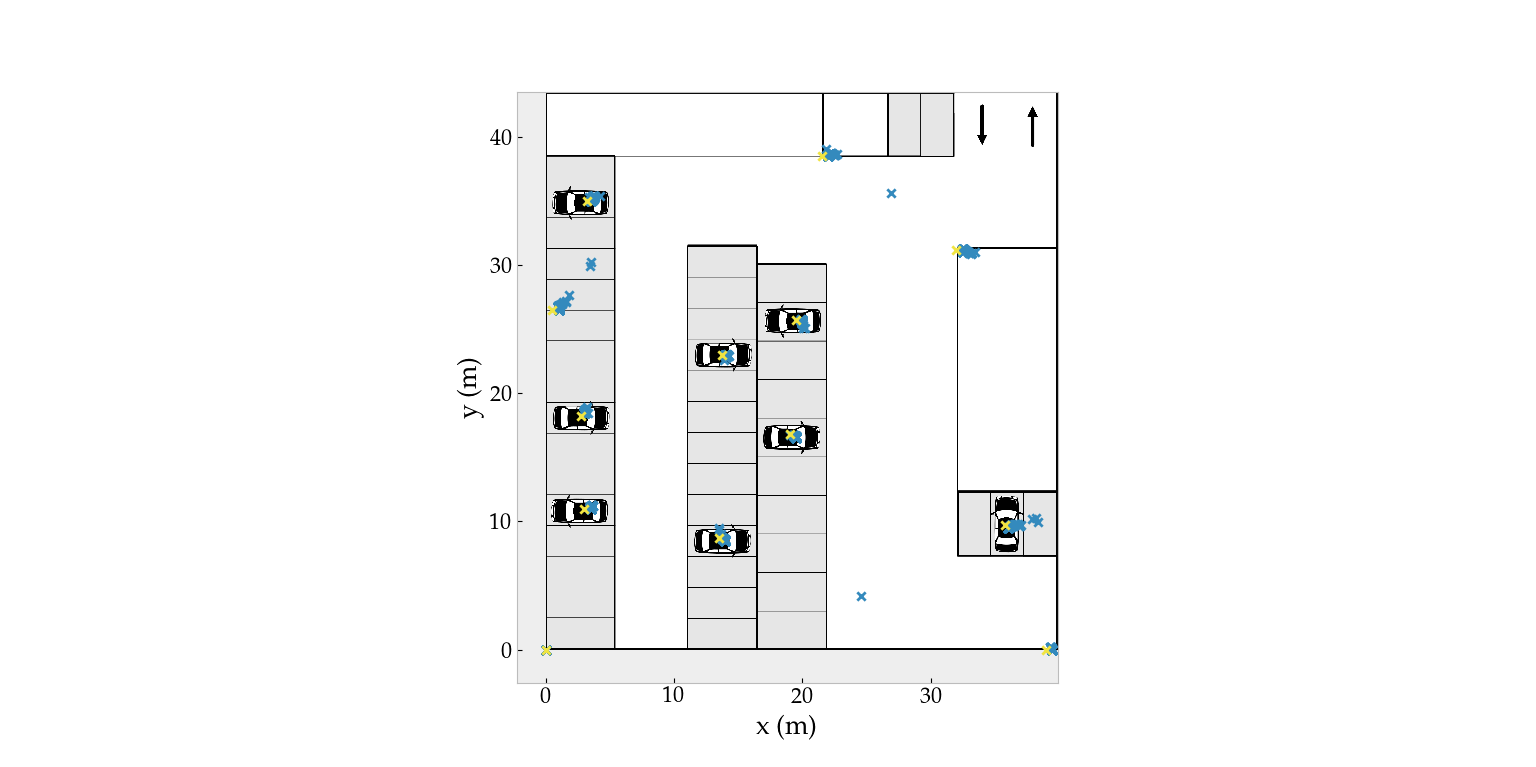}
    \caption{\RM{2}}
    \label{fig:grid_data_blue}
    \end{subfigure}
    \centering
    \begin{subfigure}[b]{0.32\textwidth}
    \centering
    \includegraphics[width=\linewidth, trim=11.5cm .8cm 12cm 2cm, clip]{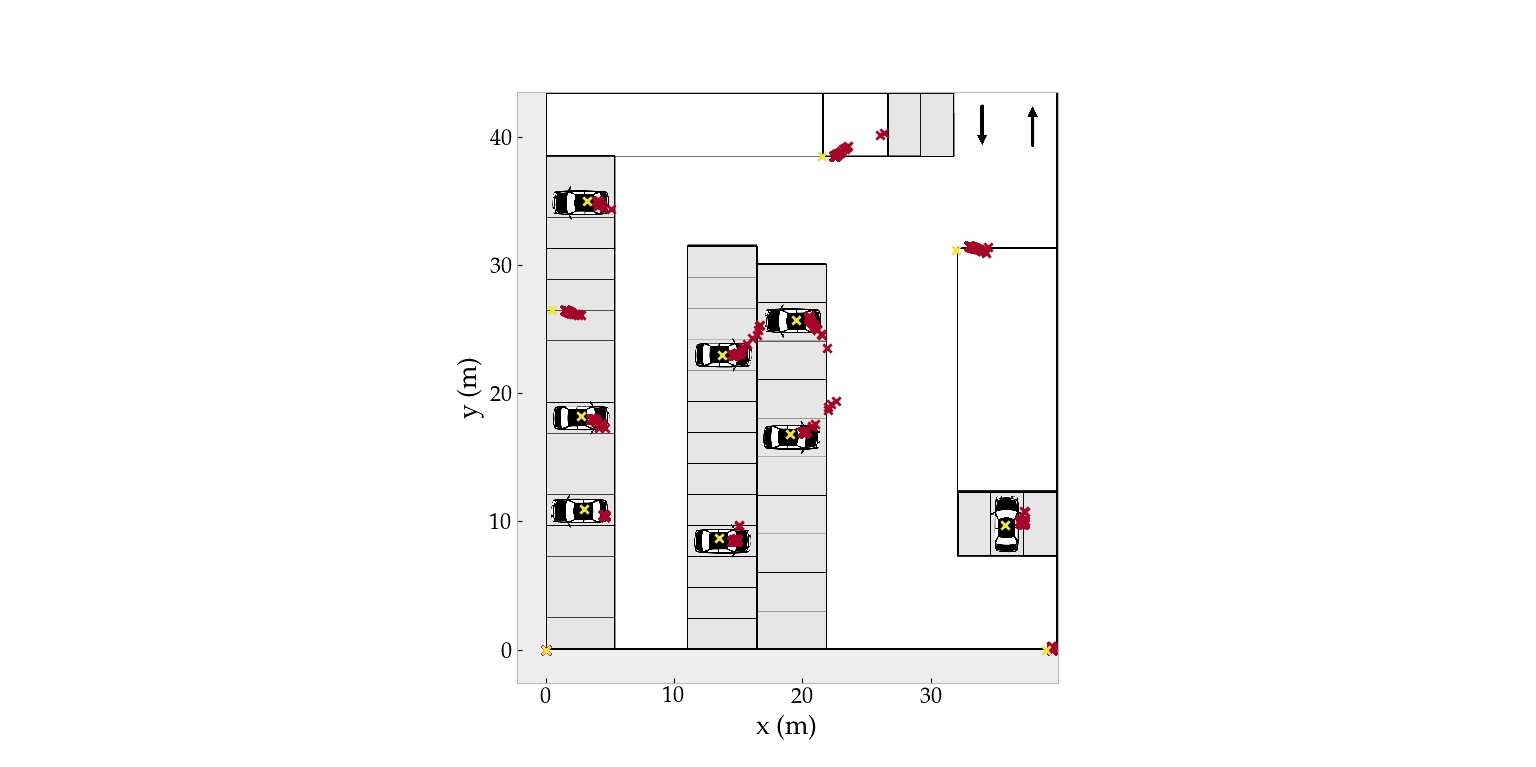}
    \caption{\RM{3}}
    \label{fig:grid_data_red}
    \end{subfigure}
    \caption{Qualitative results of CGP position estimation based on the previously presented residual PDFs (cf. \cref{fig:residual_hist}) including the references (yellow) and estimated node positions (green, blue, red).}
    \label{fig:results_pgp}
\end{figure*}

\subsection{Positioning Performance}

\begin{figure}[ht]
    \centering
    \includegraphics[width=.7\linewidth]{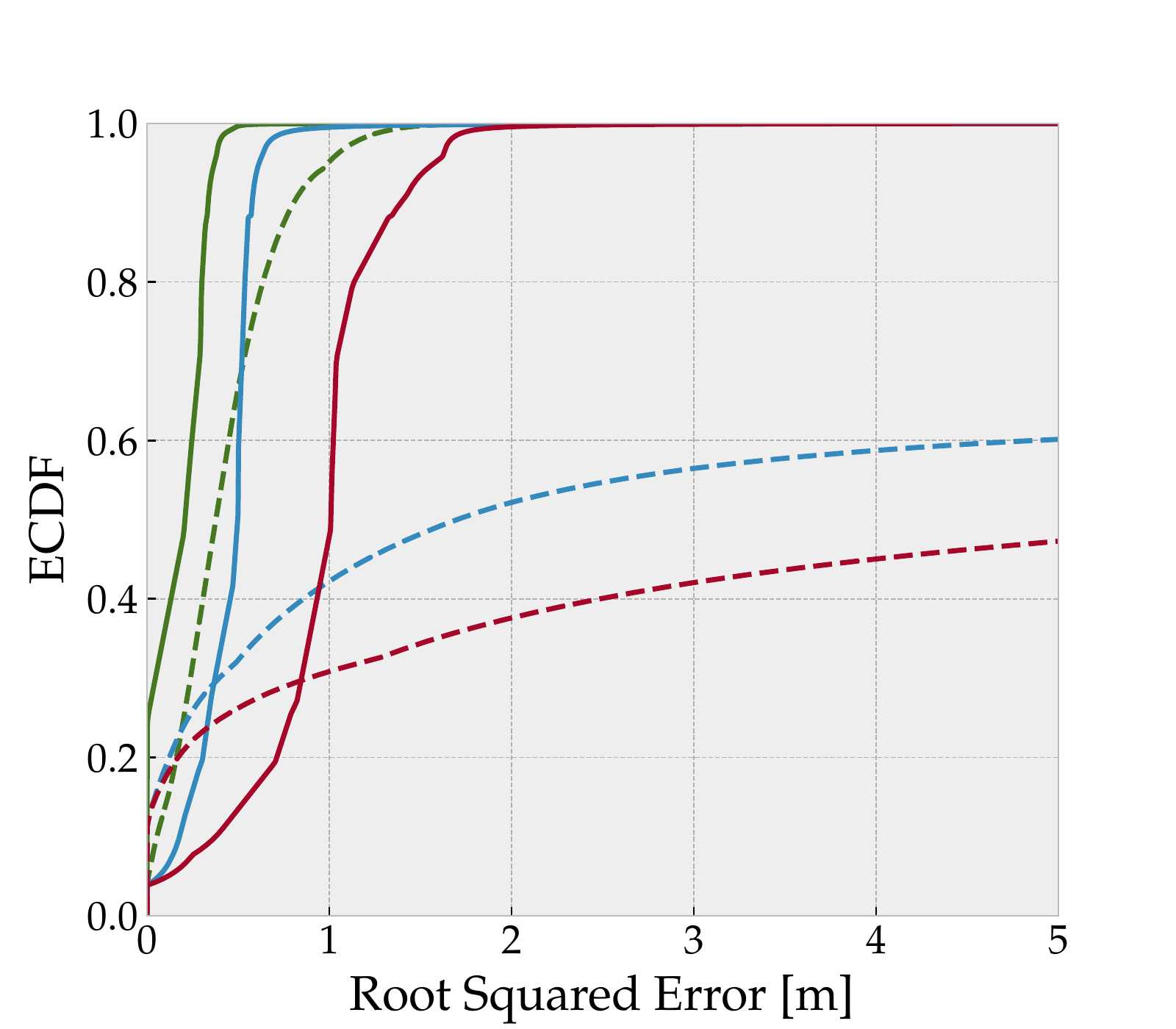}
    \caption{ECDFs for all scenarios (CGP solid, CF dotted).}
    \label{fig:ecdf}
\end{figure}

In this subsection we provide performance results of the proposed CGP approach in comparison to the aforementioned CF method and with respect to the previously introduced ranging residual distributions (cf. \cref{fig:residual_hist}). The qualitative results for each distribution are shown in \cref{fig:results_pgp}. In addition, the individual root mean square errors (RMSE) and error quantiles for both CGP and CF are detailed in \cref{tab:sim_and_results}. A graphical presentation of quantitative results is also given in \cref{fig:ecdf}, which depicts the empirical cumulative distribution functions (ECDFs) for both the CGP and the CF methods and simulation scenarios.

As expected, scenario \RM{1} achieves the most accurate position estimates. The underlying gaussian distribution contains no multipath effects, outliers and observation failures so that the ECDFs of both methods converges quickly (cf. \cref{fig:ecdf}) and reveals a RMSE of $\SI{0.41}{\meter}$ respectively $\SI{0.17}{\meter}$ for CF and CGP with a $3-\sigma$ error quantile ($99,73\%$) of $\SI{1.25}{\meter}$ and $\SI{0.44}{\meter}$. 

The results based on data sets \RM{2} and \RM{3} which entail multipath errors $\p{101}_\mathrm{mp}$ about 20\% as well as different amounts of outliers and failures, are qualitatively shown in \cref{fig:grid_data_blue,fig:grid_data_red}. Again, the proposed CGP method outperforms the baseline CF approach in terms of accuracy. This is also underlined in \cref{fig:ecdf}. 

Next to the resulting accuracy, we want to emphasize the advantages of CGP compared to CF with regards to success rate. For both scenario \RM{2} and \RM{3}, the overall CF success rate is around 0.36\%. Due to the restrictions of observation availability, the success rate drastically decreases.

For the introduced use case of location-aware smart parking applications, \RM{3} reveals that, even in the presence of 15\% outliers and 0.09\% measurement failures, a $3-\sigma$ accuracy of $\SI{1.58}{\meter}$ was achieved, which corresponds to a parking lot selective positioning accuracy.

\section{Conclusion}
\label{sec:conclusion}
In this paper, the research topic of auto-positioning for radio-based localization systems in non-static configurations was addressed. In general, auto-positioning aims to both provide position estimation of stationary anchors without time-consuming position surveying, as well as being to able to be seamlessly integrated in non-stationary network configurations. In this context, a novel approach of auto-positioning for node self-calibration, intended to provide robust state estimation in the presence of NLOS reception, ourliers and measurement failures, was presented. 

This is achieved by extending previously CF methods with a non-parametric, grid-based formulation, which we referred to as CGP. In order to emphasize the advantages of CGP in comparison with CF methods, we discussed and described three different ranging residual distributions, which correspond to different environmental and reception scenarios and are characterized by distinct error occurrence probabilities and magnitudes. 

Based on this, we empirically showed, that the proposed CGP method was able to outperform the baseline CF approach both in terms of node positioning accuracy and success rate, due to the robust formulation as well as imposing fewer requirements for network connectivity and availability of ranging measurements.

\section*{Acknowledgment}

This Project is supported by the Federal Ministry for Economic Affairs and Climate Action (BMWK) on the basis of a decision by the German Bundestag.

\bibliographystyle{IEEEtran}
\bibliography{lit}

\end{document}